\documentstyle[aps]{revtex}
\begin{document}
\def\be{\begin{equation}}
\def\ee{\end{equation}}
\def\bea{\begin{eqnarray}}
\def\eea{\end{eqnarray}}
\renewcommand{\thefootnote}{\fnsymbol{footnote}}
\def\nin{\noindent}

\twocolumn[\hsize\textwidth\columnwidth\hsize\csname@twocolumnfalse%
\endcsname

\title{Flow Equations for $U_k$ and $Z_k$}
\author{Vincenzo Branchina}
\address{Laboratoire de Physique Th\'eorique, Universit\'e Louis Pasteur,
3-5, rue de l'Universit\'e, F-67084, Strasbourg Cedex, France}

\date{\today}
\maketitle
\draft
\begin{abstract}

By considering the gradient expansion for the wilsonian effective action 
$S_k$ of a single component scalar field theory  truncated to the first two terms, 
the potential $U_k$ and the kinetic term $Z_k$, I show that the recent claim that  
different expansion of the fluctuation determinant give rise to different 
renormalization group equations for $Z_k$ is incorrect. The correct procedure to 
derive this equation is presented and the set of coupled differential equations 
for $U_k$ and $Z_k$ is definitely established.

\end{abstract} 
\pacs{pacs 11.10.Hi , 11.10.Gh \hfill}
]

During the last years there has been growing interest for the
so called Exact Renormalization Group Equation(s) (ERGE). Actually several 
questions in field theory cannot be addressed within the framework of 
perturbation theory. The entire subject of symmetry breaking, 
the problem of confinement in QCD are well known examples of
questions still waiting for an explanation.
The Wilsonian Renormalization Group Method \cite{wilson} seems to provide an 
interesting non perturbative approach to this kind of questions.
When the momentum shell of the eliminated modes 
is chosen to be infinitesimal, it results in an integro-differential equation 
for the wilsonian effective action, the action $S_k$ at the current scale
k, the Wegner-Houghton equation\cite{wegner}.
This equation is useless until a specific 
ansatz is made which allows for a systematic approximation scheme.
This can be achieved by considering the derivative expansion, whose
lowest order is the so called Local Potential Approximation (LPA).
Let us consider a single component scalar field 
theory. In the LPA $S_k$  contains only one function, the 
Local Potential $U_k(\phi)$, and the  ERGE for $S_k$ becomes a differential equation for 
$U_k$. To the next order $S_k$ contains in addition the coefficient $Z_k(\phi)$
of the lowest order derivative term $\partial_{\mu}\phi\partial_{\mu}\phi$.
While the derivation of the equation for $U_k(\phi)$ in the 
LPA is straightforward and does not present ambiguities \cite{nicol,hasen}, 
the derivation of the coupled differential equations for $U_k$ and $Z_k$ has been 
plagued by uncertaintes. 
Actually the authors of \cite{alfdar} have recently applied the expansion of the fluctuation 
determinant given in \cite{zuk} to derive these equations. They find an equation for  $Z_k$
different from the one that is obtained when the expansion of \cite{fraser} is applied.
They also computed the field anomalous dimension $\eta$. As from \cite{zuk} 
(but not from \cite{fraser}) they find the correct two loop result, they conclude that 
the expansion introduced in \cite{zuk} has the correct UV behaviour while that 
of \cite{fraser} is misleading in the UV region. 
 
By carefully reconsidering the derivation of the equations for $U_k$ and $Z_k$ 
following the method of  \cite{zuk} I show that the equation for $Z_k$ presented 
in  \cite{alfdar} is incorrect and that actually both methods \cite{zuk,fraser} 
give one and the same equation for $Z_k$. Concerning the anomalous dimension at 
two loops we note that being an $O(\hbar^2)$ result 
it comes from an infinite resummation of terms each coming from a different 
coefficient function of the gradient expansion. So the fact that $\eta$ at this order 
comes out from the two terms $U_k$ and $Z_k$ only should not be used as an 
argument to 
decide about the superiority of one expansion with respect to the other. It should rather 
been proven that the contributions coming from the infinite terms other then 
$U_k$ and $Z_k$ add up to zero. Actually we already know that this does not 
happen. By using a different but equivalent formalism, the computation of $\eta$
up to two loops order has been done in \cite{wet} where we see that the higher 
derivatives actually contribute to $\eta$. 

I will show in the following how to obtain the correct result. Before proceeding to this 
derivation I review now the functional method of \cite{zuk} that was intended to provide 
a way to compute the gradient expansion coefficients 
of the one loop effective action.
Let us  consider a single component scalar field theory. 
The effective action $\Gamma [\Phi]$ is a highly non local functional.
It can be given a quasi-local resemblance through the gradient expansion.
Up to second order in the derivatives of the field, 

\be\label{deriv}
\Gamma[\Phi]=\int d^4x \Big [ U(\Phi) +  \frac{1}{2}
Z(\Phi)\partial_\mu\Phi~\partial_\mu\Phi \Big ].
\ee

\noindent 
For definiteness we work in $d=4$ dimensions. 
In the loop expansion, on the other hand, up to one-loop order,

\be\label{split}
\Gamma[\Phi]=\Gamma_0[\Phi] + \Gamma_1[\Phi],
\ee

\nin 
where $\Gamma_0[\Phi]$ is the tree level (bare) action,~
$\Gamma_0[\Phi]=\int d^4x \Big [ U(\Phi) +  \frac{1}{2}
\partial_\mu\Phi~\partial_\mu\Phi \Big ]$ ~ and $U(\Phi)$ is the classical potential. 
$ \Gamma_1[\Phi]$ is
the one-loop contribution to $ \Gamma[\Phi]$ and can also be expanded in 
powers of the field derivatives. Again up to second order,

\be\label{oneloop}
\Gamma_1[\Phi]=\int d^4x \Big [ U_1(\Phi) +  \frac{1}{2}
Z_1(\Phi)\partial_\mu\Phi~\partial_\mu\Phi \Big ].
\ee

\nin
$U_1(\Phi)$ and $Z_1(\Phi)$ are the one-loop contributions to 
$U(\Phi)$ and to $Z(\Phi) $ respectively.

It is convenient to introduce a Dirac-like notation that will also be very useful in 
the following.
The one-loop contribution to the effective action can be written as
(from now on $U^{(n)}$ means the n-th derivative w.r.to $\Phi$)

\be\label{trace}
\Gamma_1[\Phi] = {Tr}~ ln \Bigl [ \hat P^2 +
\hat U^{(2)}(\hat\Phi(\hat x))\Bigr ].
\ee

\nin
In fact the second functional derivative of the bare action, 
$\frac{\delta^2\Gamma_0[\Phi]}{\delta\Phi(x)\delta\Phi(y)}=
\Bigl [ -\partial_x^2 + U^{(2)}(\Phi(x)) \Bigr ] \delta (x-y)$,
can be represented as the kernel of the operator
$\hat P^2 + \hat U^{(2)}(\hat\Phi(\hat x))$ in the ``$x$-representation"
once we define $\hat P_\mu$ and $\hat U^{(2)}(\hat\Phi(\hat x))$ to be 
respectively $-i\partial_\mu$ and $U^{(2)}(\Phi(x))$ in this 
representation, and introduce the notation $<x|y>= \delta (x-y)$:

\be\label{fluct}
\frac{\delta^2\Gamma_0[\Phi]}{\delta\Phi(x)\delta\Phi(y)}=
<x| \Bigl [\hat P^2 + \hat U^{(2)}(\hat\Phi(\hat x))\Bigr ]|y>.
\ee

\nin
Other representations can also be introduced. We are in particular  
interested in the ``$p$-representation", the transformation function being 
$<x|p>=\frac{1}{\sqrt V}e^{ipx}$ ($V$ is the volume), where traces are conveniently computed :

\be\label{tracy}
Tr \hat O = \sum_p <p|\hat O|p> = V \int \frac{d^4 p}{(2 \pi)^4}<p|\hat O|p>.
\ee

\nin
The notations above allow us to introduce the ``completeness relations" in the 
$x$ and $p$ representations :

\be\label{complete}
\hat I = \int d^4x |x><x| = \sum_p |p><p|. 
\ee

We can now state the method of \cite{zuk} in the following way.
First we write Eq.(\ref{trace}), a part for a meaningless infinite constant, as
(we abbreviate $\hat U^{(2)}(\hat\Phi(\hat x))$ with $\hat U^{(2)}$): 

\be\label{triv}
\Gamma_1[\Phi] = -\int_0^{\infty} du~ {Tr}~\Bigl[ \hat P^2 +
\hat U^{(2)} + u \Bigr ]^{-1}.
\ee

\nin 
Second, by the help of Eq.(\ref{tracy}), the trace in Eq.(\ref{triv}) is:

\be\label{traccia}
\sum_p <p|\Bigl [ \hat P^2 +\hat U^{(2)} + u\Bigr ]^{-1}|p>. 
\ee

\nin Third, we rewrite Eq. (\ref{traccia}) as :

\be\label{trick}
\sum_p <p|\Bigl [p^2 + \hat U^{(2)} + u -(p^2-\hat P^2)\Bigr ]^{-1}|p>.
\ee

\nin 
Eq. (\ref{trick}), were we have just added and subtracted $p^2$, contains the 
essence of the method. For any fixed value of $p$ we want to expand the operator 
$\Bigl [p^2 + \hat U^{(2)} + u -(p^2-\hat P^2)\Bigr ]^{-1}$
around $\Bigl [p^2 + \hat U^{(2)} + u \Bigr ]^{-1}$.

Up to the second order in the derivatives of the field we get
($\hat A_p = p^2 + \hat U^{(2)} + u $~ and  ~$\hat B_p = p^2-\hat P^2$)

\bea\label{expy}
{Tr}~\Bigl[& \hat P^2 + \hat U^{(2)} + u \Bigr ]^{-1} =
\sum_p <p|\Bigl [\hat A_p^{-1}~~~~~~~~~~~~~~~\nonumber\\
& + \hat A_p^{-1}\hat B_p\hat A_p^{-1}
+\hat A_p^{-1}\hat B_p\hat A_p^{-1}\hat B_p\hat A_p^{-1}\Bigl ]|p> 
\eea

\nin
The first term on the r.h.s. of Eq. (\ref{expy}), after insertion of Eq. 
(\ref{complete}) in the $x$-representation 
and integration in the $u$ variable, gives
the one loop contribution to the effective potential, 
$U_1(\Phi)=\int d^4 x \int \frac{d^4 p}{(2\pi)^4}
\ln \Bigl [p^2 + U^{(2)} (\Phi(x))\Bigr ]$. 

The second and the third term on the r.h.s. of Eq. (\ref{expy}) 
can be computed by commuting $\hat A_p^{-1}$ with $\hat B_p$ 
and applying the relation

\be\label{com}
\Bigl [\hat P_\mu, F(\hat \Phi (\hat x))\Bigr] = 
-i\frac{\partial}{\partial\hat x_\mu}F(\hat\Phi (\hat x)).
\ee
 
\nin
Inserting again Eq. (\ref{complete}) 
in the $x$-representation and integrating in the $u$ variable, 
we finally get the coefficient of $\partial_\mu\Phi(x)\partial_\mu\Phi(x)$, 
i.e. the one loop contribution $Z_1(\Phi(x))$ to $Z(\Phi(x))$. This result coincides, 
as it should, with the results of \cite{itz} and \cite{fraser}.  

Now we want to apply this method to the derivation of the 
flow equations for $U_k$ and  $Z_k$.

Let $S_k[\Phi]$ be the wilsonian effective action 
at the scale $k$. At an infinitesimal lower scale $k - \delta k$ the effective action 
$S_{k -\delta k}[\Phi]$ is given by

\bea\label{sk}
&e^{- S_{k-\delta k}[\phi]}=\int[D\eta]
e^{-S_k[\phi+\eta]}=e^{-S_k[\phi]}\times~~~~~~~~\nonumber\\ 
&\int[D\eta]
e^{-\Bigl (\int d^4 x \frac{\delta S_k[\phi]}{\delta\phi(x)}\eta(x)
 +\frac{1}{2}\int  d^4 x d^4 y \frac{\delta^2 S_k[\phi]}
{\delta\phi(x)\delta\phi(y)}\eta(x)\eta(y)\Bigr ) }.
\eea

\nin
In Eq.(\ref{sk}) we have written $\Phi(x) = \phi(x) + \eta(x)$,
separating the component $\phi(x)$ with modes form zero up to $k - \delta k$, from 
$\eta (x)$, the component with modes within the shell $[k - \delta k, k]$. 
We have also assumed that the expansion around the  background field 
$\phi(x)$ is saturated by the trivial saddle point $\eta = 0$.
In  \cite{alex} the spontaneously broken symmetry case, where 
non trivial saddle points appear, is treated. Here we limit ourselves 
to consider the unbroken case. In addition we have 
kept only terms up to $O(\eta^2)$ as in the infinitesimal shell limit 
$(\delta k \to 0)$ the gaussian approximation is exact \cite{wegner}.

Let's call ${\cal F}$ the subspace of functions 
with modes within the shell, i.e. 
${\cal F} = \Bigl \{ \psi (x) ~~~|~~~ \psi (x) = 
\sum_{\tilde p} \psi_{\tilde p} e^{i\tilde p x} ~, ~~|\tilde p|
\in [k - \delta k, k] \Bigr \}$.
The tilde  over the momentum $p$ is used from now on to indicate that 
$|\tilde p| \in [k - \delta k, k]$. 

By the help of the 
Dirac-like notation previously introduced we can write :

\bea\label{vec}
&\int d^4 x\frac{\delta S_k[\phi]}{\delta\phi(x)}\eta(x)= < s_k^{(1)} | \eta > ~~~~~~~~~~~~~~~~\nonumber\\
&\int  d^4 x d^4 y \frac{\delta^2 S_k[\phi]}
{\delta\phi(x)\delta\phi(y)}\eta(x)\eta(y) = < \eta | \hat S_k^{(2)}  | \eta >.
\eea
 
\nin
Eqs. (\ref{vec}) define the vector 
$<  s_k^{(1)} |$ and the operator $\hat S_k^{(2)}$ whose ``entries" 
in the $x$-representation are respectively 
$\frac{\delta S_k[\phi]}{\delta\phi(x)}$ and 
$\frac{\delta^2 S_k[\phi]}{\delta\phi(x)\delta\phi(y)}$.
By the help of Eqs.(\ref{vec}), 
Eq.(\ref {sk}) can be written in the compact form :
 
\be\label{ska}
e^{- S_{k-\delta k}[\phi]}= 
e^{-S_k[\phi]}\int[D\eta]~
e^{- < s_k^{(1)}| \eta >
 -\frac{1}{2} < \eta | \hat S_k^{(2)}  | \eta >}.
\ee

\nin
Note that in $< s_k^{(1)} | \eta >$ only the Fourier components  
of $| s_k^{(1)} > $ belonging to the shell $[k - \delta k, k]$ 
give contribution. Similarly for  $< \eta | \hat S_k^{(2)}  | \eta >$. 
Performing the gaussian integration in Eq. (\ref{ska}) we get :

\be\label {trln}
S_{k-\delta k}[\phi] = S_k[\phi] + 
\frac{1}{2} Tr^{'} ln \tilde S_k^{(2)} +
\frac{1}{2} < s_k^{(1)} |\Bigl[ \tilde S_k^{(2)}\Bigr]^{-1}  | s_k^{(1)} >.
\ee

\nin
Few comments are in order. It is clear that the trace has to be taken 
in the subspace ${\cal F}$, and this has been indicated by the label
${'}$ in $Tr'$.  
Moreover $\tilde S_k^{(2)}$ in Eq.(\ref{trln}) {\it is not } the operator 
$\hat S_k^{(2)}$ defined in (\ref{vec}) but rather its {\it restriction} 
to the subspace ${\cal F}$. This point has been overlooked in the previous 
literature \cite{sbjan,alfdar2,alfdar}
and Eq. (\ref{trln}) has always been written as if  $\hat S_k^{(2)}$ rather 
then $\tilde S_k^{(2)}$ appeared in it.
This illegal replacement is at the origin of the incorrect result of  \cite{alfdar}. 
  
It is easy to write down $\tilde S_k^{(2)}$. The projection operator onto 
${\cal F}$ is $\hat {\cal P} = \sum_{\tilde p} | \tilde p > < \tilde p |$
and ~$\tilde S_k^{(2)} = \hat {\cal P}\hat S_k^{(2)}\hat {\cal P}$.

As for the effective action $\Gamma(\Phi)$ in Eq.(\ref{deriv}), we write down now the 
gradient expansion for the wilsonian action $S_k$ up to the lowest order 
derivative term, i.e.

\be\label{ansatz}
{S}_k[\Phi]=
\int d^4x \Big [ U_k(\Phi) +  \frac{1}{2}
Z_k(\Phi)\partial_\mu\Phi~\partial_\mu\Phi \Big ].
\ee

From Eq.(\ref{trln}) we obtain then $U_{k-\delta k}$ and $Z_{k-\delta k}$ and finally,
sending $\delta k \to 0$, the flow equations for $U_k$ and $Z_k$.
It is not difficult to show (details will be presented in \cite{vin} )  that with the ansatz 
(\ref{ansatz}) 
$ < s_k^{(1)} |\bigl[ \tilde S_k^{(2)}\bigr]^{-1}  | s_k^{(1)} > = 0 $. 
Then in Eq. (\ref{trln}) we are only 
left with the computation of $Tr'ln \tilde S_k^{(2)}$. 

To illustrate the procedure and make our point clear, it will be sufficient to work with a 
field independent 
$Z_k$ term. For the complete treatment we only need to follow similar steps starting 
with a field dependent $Z_k$.

Writing the logarithm of the operator as
in Eq.(\ref{triv}), we have :

\be\label{sti2}
Tr'ln \tilde S_k^{(2)}= 
-\int_0^{\infty} du~ {Tr'}~\Bigl[{\cal P}\bigl( Z_k \hat P^2 + \hat U_k^{(2)} + u \bigr)
{\cal P}\Bigr]^{-1},
\ee

\nin
and expanding as in Eq.(\ref{expy}) :

\bea\label{st2}
Tr'ln \tilde S_k^{(2)}=&-\int_0^{\infty} du
\sum_{\tilde p} <\tilde p|\Bigl [\tilde A_{\tilde p}^{-1}
+ \tilde A_{\tilde p}^{-1}\tilde B_{\tilde p}\tilde A_{\tilde p}^{-1}\nonumber\\
&+\tilde A_{\tilde p}^{-1}\tilde B_{\tilde p} \hat A_{\tilde p}^{-1}\tilde B_{\tilde p}
\tilde A_{\tilde p}^{-1}\Bigl ]|\tilde p>, 
\eea

\nin
where $\tilde A_{\tilde p} ={\cal P}\hat A_{\tilde p}{\cal P}$ 
and  $\tilde B_{\tilde p} ={\cal P}\hat B_{\tilde p}{\cal P}$.

\nin
It is clear that in Eq.(\ref{st2}) only the first term gives contribution. In fact as
$\tilde B_{\tilde p}=
Z_k\sum_{\tilde p'}(\tilde p^2-\tilde p'^2) | \tilde p' > < \tilde p' |$ and 
both $\tilde p$ and $\tilde p'$ belong to the shell $[k - \delta k, k]$, the operator 
$\tilde B_{\tilde p}$ is $O(\delta k)$. From the sum over $\tilde p$ in the shell 
comes another $O(\delta k)$ and then in Eq. (\ref{st2}) all the terms apart from the 
first have to be ignored being at least $O(\delta k^2)$.

We are already in the position to compare our result with those of 
\cite{alfdar}. From Eqs.(7), (8) and (9) of that paper  we see that
the operators in the full space, i.e. $\hat A_{\tilde p}$ and $\hat B_{\tilde p}$, 
and not their restriction to the ${\cal F}$ subspace are considered. If we now replace 
in our Eq.(\ref{st2})
the operators $\tilde A_{\tilde p}$ and $\tilde B_{\tilde p}$ with 
$\hat A_{\tilde p}$ and $\hat B_{\tilde p}$ respectively, we pick up additional 
contributions from the second and third term of
Eq.(\ref{st2}). Performing for instance such a replacement in the second term, 
by the help of Eqs.(\ref{tracy}) and (\ref{complete}) we get immediately 
($a_{\tilde p}(x)= Z_k\tilde p^2 + U_k^{(2)}(\phi(x)) + u$) : 

\bea\label{spur}
&V\int\frac{d^4 \tilde p}{(2\pi)^4}
\int d^4 x<\tilde p | x > < x |\hat A_{\tilde p}^{-1} \hat B_{\tilde p}
\hat A_{\tilde p}^{-1}|\tilde p>=\nonumber\\
&-Z_k\int\frac{d^4 \tilde p}{(2\pi)^4}\int d^4 x
\frac{e^{-i\tilde p x}}{a_{\tilde p}(x)}
(\partial_\mu\partial_\mu +\tilde p^2)
\frac{e^{i\tilde p x}}{a_{\tilde p}(x)}
\eea

\nin 
From Eq.(\ref{spur}) we can easily see where the mistake originates. The operator
$\partial_\mu\partial_\mu$ acting on 
$\frac{e^{i\tilde p x}}{Z_k \tilde p^2 + U^{(2)}(\phi(x)) + u }$ 
gives rise to three terms. One is proportional to $\tilde p_\mu$
and gives zero after the angular integration in the momentum variable $\tilde p$.
Another is proportional to $-\tilde p^2$ and cancels against the $\tilde p^2$
term. Finally a third term is 

\bea\label{spu}
&\int\frac{d^4 \tilde p}{(2\pi)^4}\int d^4 x
\frac{-Z_k}{Z_k\tilde p^2 + U_k^{(2)}(\phi(x)) + u }
\partial_\mu\partial_\mu 
\frac{1}{Z_k \tilde p^2 + U_k^{(2)}(\phi(x)) + u }\\
& ~~\nonumber
\eea

\nin
This term is not zero and gives additional spurious contributions to $Z_{k - \delta k}$.

Let's now give a closer look to the first term of Eq.(\ref{st2}):
\be\label{only}
\sum_{\tilde p} <\tilde p|\Bigl[{\cal P}( Z_k \tilde p^2 + \hat U_k^{(2)} + u ){\cal P}
\Bigr ]^{-1}|\tilde p >.
\ee

\nin
In  \cite{alfdar} this contribution is written as:

\be\label{refy}
\int\frac{d^4 \tilde p}{(2\pi)^4}\int d^4x
\frac{1}{Z_k\tilde p^2 + U_k^{(2)}(\phi(x)) + u }
\ee 

\nin
Again this would be right if we could ignore the presence
of the projection operator ${\cal P}$ in Eq.(\ref{only}). As 
${\cal P}|\tilde p > = |\tilde p >$, this would amount to replace the inverse 
$\tilde A_{\tilde p}^{-1}$in ${\cal F}$
of the restricted operator $\tilde A_{\tilde p}$ with the restriction 
in ${\cal F}$ of the inverse operator, ${\cal P}\hat A_{\tilde p}^{-1}{\cal P}$ . 
This replacement would be correct if the operator $\hat A_{\tilde p}$ was diagonal
in the $p$-representation. As we see from its definition, 
this is certainly not the case. 

Actually the projection operator in Eq.(\ref{only}) is not easy to handle unless we 
develop  
$[{\cal P}( Z_k \tilde p^2 + \hat U_k^{(2)}(\hat\phi(\hat x)) + u ){\cal P}]^{-1}$ around the 
{\it diagonal} operator
$[{\cal P}( Z_k \tilde p^2 + U_k^{(2)}(\phi_0) + u ){\cal P}]^{-1}$, obtained expanding 
$\phi(x)$ around the constant value  $\phi_0$ ($U_k^{(n)}(\phi_0)= U_{k0}^{(n)}$):

\be\label{phio}
U_k^{(2)}(\phi) = U_{k0}^{(2)} + U_{k0}^{(3)}\partial_\mu\phi +
\frac{1}{2}U_{k0}^{(4)}\partial_\mu\phi\partial_\mu\phi +\cdots.
\ee

\nin
If we now insert Eq.(\ref{phio}) in Eq.(\ref{only}) and then integrate in the $u$ variable,
we get the same result we would have obtained if we had started by expanding the logarithm 
in the fluctuation determinant of Eq. (\ref{trln}) around 

\be\label{loga}
ln [{\cal P}(Z_k\hat P^2 + U_{k0}^{(2)}){\cal P}]
\ee

\nin
i.e. as if we had used from the very beginning the expansion of \cite{fraser}.

We can now come to our reassuring conclusion.
There is no contradiction between the two methods of
\cite{zuk} and \cite{fraser} that, when applied to the fluctuation determinant in 
Eq.(\ref{trln}), give {\it one and the same} result 
for $Z_{k-\delta k}$, i.e. the same flow equation for $Z_k$. We have also learnt that, 
due to the constraint imposed by the presence of the projection operator ${\cal P}$ 
the method of \cite{zuk} trivially turns to the method of \cite{fraser}. 

I give now the couple of differential equations for $U_k$ and 
$Z_k$ that are obtained once the full $\phi$ dependence of $Z_k$ is taken into account
($A= Z_k k^2 + U_k^{(2)}$ and $Z_k^{(n)}$, $A^{(n)}$ are derivatives w.r. to the field):

\be\label{poteqz}
k\frac{\partial}{\partial k}U_k=-
\frac{k^4}{ 16 \pi^2}ln A ~~~~~~~~~~~~~~~~~~~~~~~~~~~~~~~~~~~~
\ee

\bea\label{zfras}
k\frac{\partial}{\partial k}Z_k=&-\frac{k^4 }{16 \pi^2}\Bigl (
\frac{Z_k^{(2)}}{A}-
\frac{2 Z_k^{(1)} A^{(1)}}{A^2}-
\frac{{Z_k^{(1)}}^2 k^2}{ 4 A^2}+\nonumber\\
&\frac{{Z}_k {A^{(1)}}^2}{ A^3}+
\frac{Z_k^{(1)} Z_k A^{(1)} k^2}{ A^3}-
\frac{Z_k^2 {A^{(1)}}^2 k^2}{A^4} \Bigr )
\eea

These equations have already been presented in \cite{alextesi,bona} and 
similar equations in \cite{sbjan,alfdar2},
but a word of caution has to be said concerning their derivation.
In \cite{sbjan,alfdar2} where to derive the flow equations the method of 
\cite{fraser} was applied, the presence of the projection operator ${\cal P}$ 
was not taken into account. In addition there is one point that I have deliberately 
avoided to mention up to now. The presence of the projection operator ${\cal P}$, 
that is due to the choice of a sharp cut-off for the mode elimination,
has another effect: it potentially brings additional terms in the derivation of 
$Z_{k-\delta k}$ from Eq.(\ref{trln}). Again due to the neglect of ${\cal P}$,
these terms went unnoticed in \cite{sbjan,alfdar2}. These derivations are then not
on a firm foot.  
In \cite{alextesi} Eq.(\ref{zfras}) is obtained by considering the 
fluctuation operator directly in the p-representation. No ambiguity 
is then  present concerning the restriction of the operator $\hat S_k^{(2)}$. 
Nevertheless the method employed, namely the choice of a particular 
background field $\phi(x)$, carries ambiguities
due to the appearance of these additional non-analytic terms \cite{sak} in  
contradiction with the gradient expansion itself. Those terms are there
neglected without any justification and the whole method seems not to be 
firmly established. 

In \cite{bona} the first step beyond 
the LPA is taken by brute force integration in the space of the 
Fourier components of the fluctuation field, i.e. without any reference to the 
functional methods of \cite{zuk} and \cite{fraser}. Again a special non constant background 
field is chosen to extract the differential equation for $Z_k$, namely a field with a
single Fourier component  
$\phi(x)\sim [\varphi_q e^{iqx} + \varphi_{-q} e^{-iqx}]$, with 
$q\sim 0$. As in  \cite{alextesi} additional terms non analytic in 
$q^2$ are found but, for the first time, the method allowed to 
compare the magnitude of these terms with the ones that are retained 
in establishing equation (\ref{zfras}). In this way it was possible to find 
the conditions under which these terms can be safely neglected, i.e. the 
validity conditions of Eq.(\ref {zfras}).
But at that time it was not yet clear that the result of \cite{alfdar} was incorrect, i.e.
if Eq.(\ref{zfras}) or the corresponding  Eq.(18) of \cite{alfdar} was the correct one.
Actually in \cite{bona} we incorrectly argued that both equations could be right 
as being different approximations for different physical situations. Only now it appears 
clearly that the system of Eqs.(\ref {poteqz}) and (\ref {zfras}) 
{\it is} the next order of approximation to the Wegner-Houghton equation in the gradient 
expansion, after the LPA of \cite{nicol,hasen}.

In a forthcoming paper \cite{vin} I will 
present a complete analysis of the problems related to the presence of a sharp cut-off. 
It is sufficient to say here that the 
conclusions of this work, more general than that of \cite{bona} as no reference to a specific 
background field is done, actually meet in this respect those of \cite{bona}: for a 
sufficiently 
smooth background field, these additional terms can be neglected and Eqs.(\ref {poteqz})
and (\ref {zfras}) give the correct approximation to the Wegner-Houghton equation at this 
order. 

\vskip 10 pt
I would like to thank H. Mohrbach for useful discussions.

\end{document}